\documentclass[twocolumn,amsmath]{aastex702}
\journalinfo{The Astrophysical Journal, 1012, 121 (2026)}

\begin{document}

\shorttitle{Polarization of FRB~20240209A}
\shortauthors{Zhang et al.}

\title{Polarization and Burst Properties of the Repeating FRB~20240209A in a Quiescent Elliptical Host}

\author[orcid=0000-0002-8744-3546]{Yong-Kun Zhang}
\affiliation{State Key Laboratory of Radio Astronomy and Technology, National Astronomical Observatories, Chinese Academy of Sciences, Beijing 100101, People's Republic of China}
\email[show]{ykzhang@nao.cas.cn}

\author[orcid=0000-0002-0475-7479]{Yi Feng}
\affiliation{Research Center for Astronomical Computing, Zhejiang Laboratory, Hangzhou 311100, People's Republic of China}
\affiliation{Institute for Astronomy, School of Physics, Zhejiang University, Hangzhou 310027, People's Republic of China}
\email[show]{yifeng@zhejianglab.org}

\author[orcid=0000-0003-3010-7661]{Di Li}
\affiliation{New Cornerstone Science Laboratory, Department of Astronomy, Tsinghua University, Beijing 100084, People's Republic of China}
\affiliation{State Key Laboratory of Radio Astronomy and Technology, National Astronomical Observatories, Chinese Academy of Sciences, Beijing 100101, People's Republic of China}
\email[show]{dili@mail.tsinghua.edu.cn}

\author[orcid=0000-0001-5649-2591]{Jin-Tao Xie}
\affiliation{School of Computer Science and Engineering, Sichuan University of Science and Engineering, Yibin 644000, People's Republic of China}
\email{xiejintao@suse.edu.cn}

\author[orcid=0000-0002-9390-9672]{Chao-Wei Tsai}
\affiliation{State Key Laboratory of Radio Astronomy and Technology, National Astronomical Observatories, Chinese Academy of Sciences, Beijing 100101, People's Republic of China}
\affiliation{Institute for Frontiers in Astronomy and Astrophysics, Beijing Normal University, Beijing 102206, People's Republic of China}
\affiliation{University of Chinese Academy of Sciences, Beijing 100049, People's Republic of China}
\email{cwtsai@nao.cas.cn}

\author[orcid=0000-0002-3386-7159]{Pei Wang}
\affiliation{State Key Laboratory of Radio Astronomy and Technology, National Astronomical Observatories, Chinese Academy of Sciences, Beijing 100101, People's Republic of China}
\affiliation{Institute for Frontiers in Astronomy and Astrophysics, Beijing Normal University, Beijing 102206, People's Republic of China}
\email{wangpei@nao.cas.cn}

\author{Meng-Yao Xue}
\affiliation{State Key Laboratory of Radio Astronomy and Technology, National Astronomical Observatories, Chinese Academy of Sciences, Beijing 100101, People's Republic of China}
\email{mengyaoxue@nao.cas.cn}

\makeatletter
\gdef\@@email{\parbox[t]{0.42\textwidth}{\raggedright ykzhang@nao.cas.cn, yifeng@zhejianglab.org, dili@mail.tsinghua.edu.cn}}
\makeatother

\begin{abstract}
    We report polarimetric observations of the repeating fast radio burst (FRB) source FRB~20240209A with the Green Bank Telescope at L band and 800 MHz. This source has previously been associated with a quiescent elliptical galaxy with a mass-weighted stellar age of approximately 11~Gyr. We detected three bursts, and the brightest burst resolves into eight components that drift toward lower frequencies. The bursts have mutually consistent rotation measures (RMs), and the small extragalactic RM agrees with the low RM predicted from the small host-galaxy dispersion measure (\({\rm DM}_{\rm host}\)). The linear polarization fraction decreases toward lower frequencies within the brightest burst, consistent with depolarization by an RM scatter of \(\sigma_{\rm RM}\simeq3.5~{\rm rad~m^{-2}}\). This RM scatter is comparable to those of several repeaters in star-forming galaxies. Therefore, FRBs in both star-forming and old, quiescent hosts can reside in relatively complex magnetoionic environments, which suggests that FRB sources may form through multiple channels.
\end{abstract}


\section{Introduction}
\label{sec:intro}

Fast radio bursts (FRBs) are bright pulses of coherent radio emission from extragalactic sources \citep{2007Sci...318..777L,2013Sci...341...53T,2022A&ARv..30....2P}, with durations from microseconds to milliseconds \citep{2023NatAs...7.1486S}. Some FRBs repeat, showing that their engines can remain active over many emission episodes \citep{2016Natur.531..202S}. The FRB-like burst from the Galactic magnetar SGR~1935+2154 demonstrated that magnetars can generate such emission \citep{2020Natur.587...54C,2020Natur.587...59B}. The formation pathways and possible diversity of extragalactic FRB engines remain uncertain. Comparisons between the burst-sequence statistics of repeating FRBs and those of magnetar flares and other energetic transients constrain FRB triggering mechanisms \citep{2024SciBu..69.1020Z,2026SCPMA..6909511X}. Host-galaxy properties constrain the stellar populations from which FRB sources arise, and propagation effects probe the plasma surrounding active sources.

Several observations connect FRBs to young stellar environments. FRB~20180916B lies near a young star-forming complex, and FRB~20190520B resides in a star-forming dwarf galaxy \citep{2021ApJ...908L..12T,2022Natur.606..873N,2026SciBu..71...76N}. Among 23 FRB host galaxies with secure associations, 20 were classified as globally star forming \citep{2023ApJ...954...80G}. These observations are consistent with a formation channel connected to massive-star evolution. The DSA-110 host sample contains fewer low-mass galaxies than expected from a star formation rate (SFR)--weighted galaxy population, suggesting an additional dependence of FRB production on host properties such as stellar mass or metallicity \citep{2024Natur.635...61S}. Gas-phase metallicity measurements of FRB hosts show no metallicity preference \citep{2026ApJ...997L...6Y}. FRB~20200120E is located in an old globular cluster (GC) in M81 and requires a formation pathway compatible with an old stellar population \citep{2022Natur.602..585K}. Based on this source, old stellar populations have been estimated to contribute at least a few percent of the FRB rate, and a comparable fraction of localized FRBs has been predicted to lie in elliptical galaxies or at large host offsets \citep{2022MNRAS.510.1867L}.

Polarimetry provides a direct probe of the magnetized plasma along the sight lines toward FRBs as well as pulsars \citep{2022A&ARv..30....2P,2025SciBu..70.1568Z}. The rotation measure (RM) is proportional to the path integral of the electron density and the line-of-sight magnetic field. FRB~20121102A has an exceptionally large RM that declined over several years \citep{2018Natur.553..182M,2021ApJ...908L..10H}, and the RM of FRB~20190520B changed sign twice within 17 months \citep{2023Sci...380..599A}. Monitoring with FAST and CHIME/FRB has found RM variations in a substantial fraction of repeaters \citep{2025SCPMA..6889511F,2025ApJ...982..154N}. Frequency-dependent depolarization provides a complementary measure of small-scale structure. In the RM-scattering model \citep{1966MNRAS.133...67B}, unresolved propagation paths with different Faraday depths reduce the observed linear polarization toward lower frequencies, and \(\sigma_{\rm RM}\) quantifies the RM spread among those paths \citep{2022Sci...375.1266F,2022ApJ...928L..16Y}. In the sample of \citet{2022Sci...375.1266F}, the two repeaters with compact persistent radio sources (PRSs), FRBs~20121102A and 20190520B, have the largest \(\sigma_{\rm RM}\). This association has been interpreted as evidence for dense, turbulent plasma in young environments \citep{2022Sci...375.1266F,2022ApJ...928L..16Y}. The small \(\sigma_{\rm RM}\) of FRB~20180916B has been associated with a less turbulent environment \citep{2022Sci...375.1266F}. RM and \(\sigma_{\rm RM}\) therefore describe the circumburst medium on scales that host-galaxy observations cannot resolve.

FRB~20240209A is an active repeater discovered with CHIME/FRB and localized with the CHIME--KKO baseline to the outskirts of a luminous galaxy \citep{2025ApJ...979L..21S}. Its host at \(z=0.1384\) is a massive, quiescent elliptical galaxy with a mass-weighted stellar age of approximately \(11\,{\rm Gyr}\) \citep{2025ApJ...979L..22E}. This association identified the first confirmed elliptical FRB host and the first quiescent host of a repeating FRB \citep{2025ApJ...979L..21S,2025ApJ...979L..22E}. The source has a projected offset of \(40\pm5\,{\rm kpc}\) from the host center, and a globular-cluster origin remains plausible \citep{2025ApJ...979L..21S}. The dispersion-measure (DM) budget leaves little room for host and local contributions. \citet{2025ApJ...979L..21S} therefore inferred a clean local environment and predicted a low RM, and they deferred the polarization analysis to future work.

Here, we report Green Bank Telescope (GBT) polarimetric observations of FRB~20240209A with the 800~MHz and L-band receivers. These observations test whether an active repeater associated with an old, quiescent host shows the magnetoionic signatures seen in repeaters in star-forming galaxies. Section~\ref{sec:data} describes the observations and data processing. Section~\ref{sec:results} presents the burst properties, polarization measurements, and a search for submillisecond component spacing. Section~\ref{sec:discussion} discusses the burst substructure and the magnetoionic environment. Section~\ref{sec:conclusion} summarizes our conclusions.

\section{Observations and data processing}
\label{sec:data}

We observed FRB~20240209A on 2024 August 18 with the GBT using the Versatile GBT Astronomical Spectrometer (VEGAS) back end under project GBT24A-484. The observations comprised two consecutive 2 hr sessions, using the L-band receiver from 13:30 to 15:30 UTC and the 800 MHz receiver from 15:30 to 17:30 UTC. The data were coherently dedispersed in real time at a DM of \(176.57~{\rm pc~cm^{-3}}\), as reported in \citet{2024ATel16670....1S}.

The L-band data were recorded at a central frequency of \(1500.4~{\rm MHz}\), with \(800~{\rm MHz}\) bandwidth, 1024 frequency channels, and a time resolution of \(20.48~\mu{\rm s}\). The usable frequency range was \(1100.8\)--\(1900.0~{\rm MHz}\). The 800 MHz data were recorded at a central frequency of \(820.2~{\rm MHz}\), with \(200~{\rm MHz}\) bandwidth, 512 frequency channels, and a time resolution of \(40.96~\mu{\rm s}\), covering \(720.4\)--\(920.0~{\rm MHz}\).

We searched the total-intensity data with DRAFTS\footnote{\url{https://github.com/SukiYume/DRAFTS}}, a deep-learning-based single-pulse search pipeline \citep{2025ApJS..276...20Z}. The raw data were first dedispersed at \({\rm DM}=176.57~{\rm pc~cm^{-3}}\). We then downsampled the data in time by a factor of 8 for the L-band data and 4 for the 800 MHz data and divided them into segments of 512 time samples. Each segment was resized to \(512\times512\) pixels and passed to the DRAFTS classification model. We inspected all candidates with a classification probability above 50\% and rejected those attributable to radio-frequency interference (RFI).

We calibrated the data with PSRCHIVE \citep{2004PASA...21..302H}. Flux calibration used 3C196 on/off scans obtained with the corresponding receiver setup, and polarization calibration used the nearest noise-diode scan from the same receiver. We then formed flux- and polarization-calibrated Stokes \(IQUV\) dynamic spectra, subtracted the off-burst baseline in each frequency channel, and masked RFI channels.

For each burst, we formed a frequency-averaged Stokes-\(I\) profile over the channels containing burst emission, which we labeled by hand after masking RFI channels. The burst window is the contiguous span in which this profile exceeds 3 times the off-pulse rms noise. The peak flux density \(S_{\rm pk}\) is the maximum of the profile, and the fluence \(\mathcal{F}\) is the integral of the profile over the burst window. The fluence was measured over the observed burst bandwidth \(\Delta\nu_{\rm obs}\). The equivalent width is
\begin{equation}
    W_{\rm eq}=\frac{\mathcal{F}}{S_{\rm pk}},
    \label{eq:equivalent_width}
\end{equation}
which is the width of a top-hat pulse with the same peak flux density and fluence and is shorter than the burst window.

The isotropic-equivalent radio energy was calculated using
\begin{equation}
    E_{\rm iso}=10^{39}{\rm erg}\,
    \frac{4\pi}{1+z}
    \left(\frac{D_{\rm L}}{10^{28}{\rm cm}}\right)^2
    \left(\frac{\mathcal{F}}{{\rm Jy\,ms}}\right)
    \left(\frac{\Delta\nu_{\rm obs}}{{\rm GHz}}\right),
    \label{eq:energy}
\end{equation}
where \(D_{\rm L}=675.0~{\rm Mpc}\) is the luminosity distance at the host-galaxy redshift \(z=0.1384\) for the Planck18 cosmology \citep{2020A&A...641A...6P}.

The RM in the rest frame of the plasma is
\begin{equation}
    {\rm RM}=0.812\int_{\rm source}^{\rm observer}n_e B_\parallel\,dl~{\rm rad~m^{-2}},
    \label{eq:rm}
\end{equation}
where \(n_e\) is the electron density in \({\rm cm^{-3}}\), \(B_\parallel\) is the line-of-sight magnetic field in \(\mu{\rm G}\), and the path length \(dl\) is in pc. We measured the RM with RM synthesis by derotating Stokes \(Q\) and \(U\) over a grid of trial RM values and selecting the value that maximized the summed linear polarization across the burst window \citep{2005A&A...441.1217B}. After correcting for the best-fit RM, we calculated the measured linear polarization in each time sample as \(L_i=(Q_i^2+U_i^2)^{1/2}\) and applied the positive-bias correction \citep{2001ApJ...553..341E}
\begin{equation}
    L_{{\rm de\mbox{-}bias},i}=
    \begin{cases}
        \sigma_I\sqrt{\left(L_i/\sigma_I\right)^2-1}, & L_i/\sigma_I>1.57,\\
        0, & L_i/\sigma_I\leq1.57,
    \end{cases}
    \label{eq:linear_debias}
\end{equation}
where \(\sigma_I\) is the rms noise of the Stokes-\(I\) profile. The linear and circular polarization fractions were calculated over the burst window as
\begin{equation}
    \frac{L}{I}=\frac{\sum_i L_{{\rm de\mbox{-}bias},i}}{\sum_i I_i},\qquad
    \frac{V}{I}=\frac{\sum_i V_i}{\sum_i I_i}.
    \label{eq:polarization_fractions}
\end{equation}
The polarization position angle was calculated as
\begin{equation}
    {\rm PA}_i=\frac{1}{2}\operatorname{atan2}(U_i,Q_i),
    \label{eq:pa}
\end{equation}
for samples with \(L_i/\sigma_I>1.57\).

\section{Results}
\label{sec:results}

\subsection{Burst properties}
\label{subsec:burst_properties}

\begin{figure*}[htb!]
    \centering
    \includegraphics[width=\textwidth]{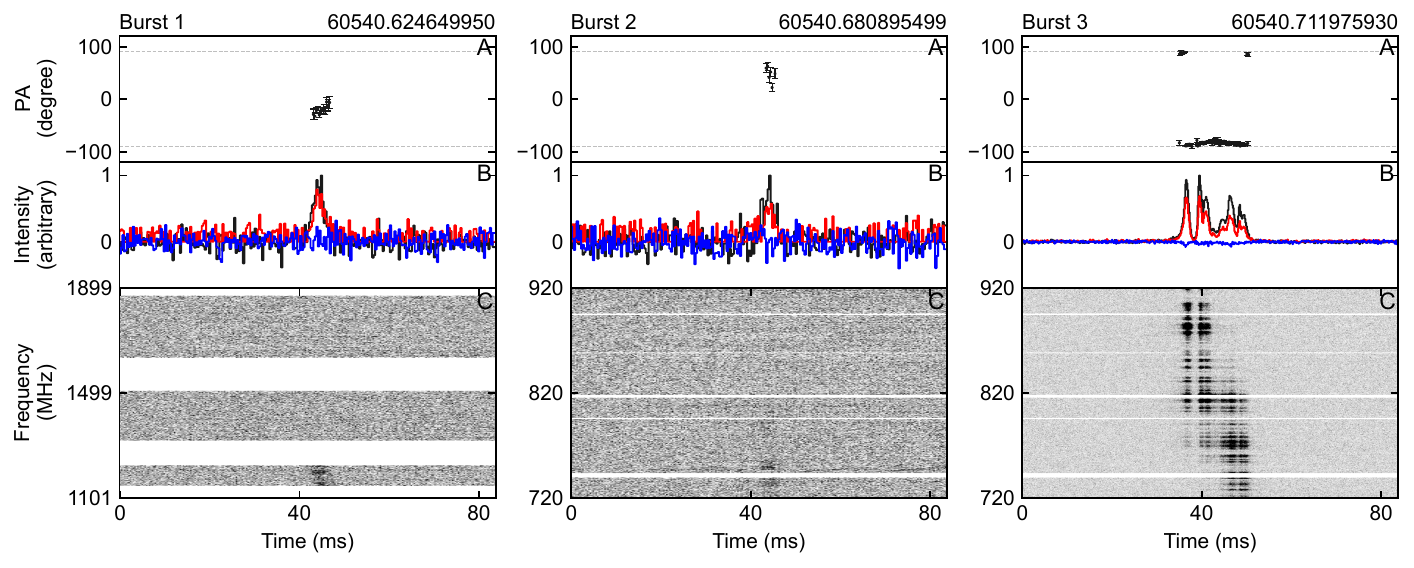}
    \caption{Polarization properties of the three bursts from FRB~20240209A observed with the GBT. The three columns show Bursts 1, 2, and 3, respectively. \textbf{(A)} RM-corrected polarization position angle (PA). The gray dashed lines mark \(\pm90^\circ\). \textbf{(B)} Frequency-integrated profiles of total intensity (black), linear polarization (red), and circular polarization (blue). \textbf{(C)} Dedispersed Stokes-\(I\) dynamic spectra. The white bands indicate masked RFI channels.}
    \label{fig:bursts}
\end{figure*}

We detected three bursts from FRB~20240209A, one with the L-band receiver and two with the 800 MHz receiver. Figure~\ref{fig:bursts} shows their dynamic spectra and polarization profiles, and Table~\ref{tab:burst_properties} lists their measured properties. Bursts 1 and 2 have band-limited fluences of \(1.41\pm0.07\) and \(4.1\pm0.3~{\rm Jy~ms}\), respectively. Burst 3 is much brighter, with an integrated signal-to-noise ratio (S/N) of 309, a fluence of \(51.7\pm0.2~{\rm Jy~ms}\), and an isotropic-equivalent energy of \((4.69\pm0.02)\times10^{39}~{\rm erg}\).

We optimized the DM of Burst 3 with \texttt{DM\_phase} and obtained \({\rm DM}=174.93\pm0.12~{\rm pc~cm^{-3}}\) \citep{2019ascl.soft10004S}. At this DM, the burst profile resolves into eight components whose emission shifts toward lower frequencies at later arrival times. This ``sad-trombone'' pattern is widely observed in repeating FRBs \citep{2019ApJ...876L..23H,2021ApJ...923....1P,2022RAA....22l4001Z,2023MNRAS.519..666J}. To measure the drift, we divided the dynamic spectrum into eight time intervals with boundaries midway between adjacent component peaks. Within each interval, we averaged the dynamic spectrum over time and calculated the intensity-weighted mean frequency of the resulting spectrum. A linear fit of these eight frequency centroids against the component peak times gives \(\dot{\nu}=-8.5\pm0.9~{\rm MHz~ms^{-1}}\).

The drift rate of Burst~3 falls within the ranges reported for FRBs~20180916B and 20220912A, approximately \(-36\) to \(-8~{\rm MHz~ms^{-1}}\), while FRB~20121102A shows a characteristic rate of about \(-200~{\rm MHz~ms^{-1}}\) at 1.1--1.7 GHz \citep{2019ApJ...876L..23H,2022ApJ...932...98S,2023MNRAS.526.2039H}. These comparisons involve different observing bands and drift estimators. Downward drift can arise through magnetospheric radius-to-frequency mapping or through a decreasing characteristic emission frequency as a relativistic shock decelerates \citep{2022ApJ...927..105W,2019MNRAS.485.4091M}. The drift of Burst~3 is qualitatively compatible with both model classes.

\begin{figure*}[htb!]
    \centering
    \includegraphics[width=0.85\textwidth]{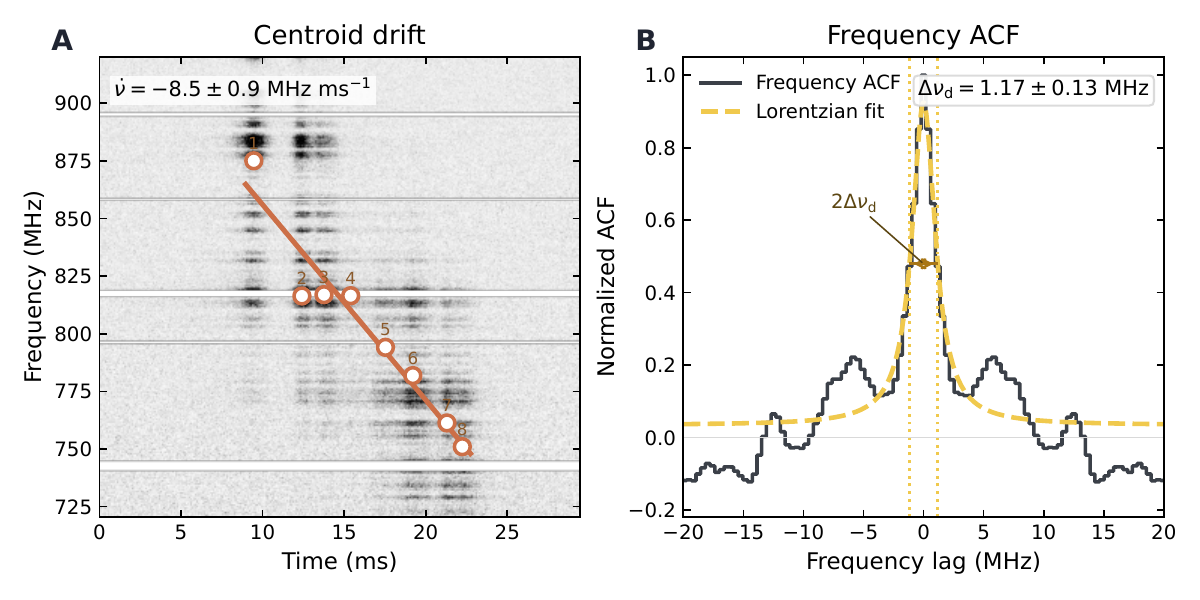}
    \caption{Frequency drift and spectral autocorrelation of Burst 3. \textbf{(A)} Dedispersed Stokes-\(I\) dynamic spectrum. The orange circles mark the frequency centroids, and the orange line represents the linear fit. \textbf{(B)} Normalized frequency autocorrelation function (ACF) shown as a gray curve, together with its Lorentzian fit shown as a yellow dashed line. The yellow dotted lines mark \(\pm\Delta\nu_{\rm d}\).}
    \label{fig:drift_scint}
\end{figure*}

To measure the scintillation bandwidth of Burst 3, we integrated the dynamic spectrum over the burst window and calculated the frequency autocorrelation function (ACF) of the resulting spectrum. A Lorentzian fit to the ACF gives \(\Delta\nu_{\rm d}=1.17\pm0.13~{\rm MHz}\) at 820 MHz (Figure~\ref{fig:drift_scint}).

\begin{splitdeluxetable*}{lccccccBlcccccccc}
    \tabletypesize{\scriptsize}
    \tablewidth{0pt}
    \tablecaption{The Properties of the Three FRB~20240209A Bursts Observed with the GBT\label{tab:burst_properties}}
    \tablehead{
        \colhead{Burst} & \colhead{MJD} & \colhead{UTC} & \colhead{Band} & \colhead{S/N} & \colhead{$W_{\rm eq}$} & \colhead{$S_{\rm pk}$} & \colhead{Burst} & \colhead{$\mathcal{F}$} & \colhead{$\Delta\nu_{\rm obs}$} & \colhead{$E_{\rm iso}$} & \colhead{RM} & \colhead{$L/I$} & \colhead{$V/I$} & \colhead{$\dot{\nu}$} & \colhead{$\Delta\nu_{\rm d}$} \\
        \colhead{} & \colhead{} & \colhead{} & \colhead{} & \colhead{} & \colhead{(ms)} & \colhead{(Jy)} & \colhead{} & \colhead{(Jy ms)} & \colhead{(MHz)} & \colhead{(erg)} & \colhead{(rad m$^{-2}$)} & \colhead{(\%)} & \colhead{(\%)} & \colhead{(MHz ms$^{-1}$)} & \colhead{(MHz)}
    }
    \startdata
        1 & 60540.624649950 & 2024-08-18T14:59:29.756 & L & 23.7 & $2.57\pm0.20$ & $0.55\pm0.03$ & 1 & $1.41\pm0.07$ & 78 & $(5.3\pm0.3)\times10^{37}$ & $-40\pm8$ & $80\pm6$ & $8\pm5$ & \nodata & \nodata \\
        2 & 60540.680895499 & 2024-08-18T16:20:29.371 & 800 MHz & 14.0 & $2.34\pm0.19$ & $1.74\pm0.09$ & 2 & $4.1\pm0.3$ & 32 & $(6.2\pm0.4)\times10^{37}$ & $-60^{+15}_{-18}$ & $60\pm8$ & $3\pm7$ & \nodata & \nodata \\
        3 & 60540.711975930 & 2024-08-18T17:05:14.720 & 800 MHz & 309 & $5.98\pm0.02$ & $8.65\pm0.02$ & 3 & $51.7\pm0.2$ & 190 & $(4.69\pm0.02)\times10^{39}$ & $-45.5\pm0.2$ & $63.6\pm0.4$ & $-5.5\pm0.3$ & $-8.5\pm0.9$ & $1.17\pm0.13$ \\
    \enddata
    \tablecomments{$W_{\rm eq}$, $S_{\rm pk}$, $\mathcal{F}$, and $E_{\rm iso}$ denote the equivalent width, peak flux density, fluence, and isotropic-equivalent energy, respectively. The quantity $\Delta\nu_{\rm obs}$ is the observed burst bandwidth. The fluence $\mathcal{F}$ is measured over $\Delta\nu_{\rm obs}$. The isotropic-equivalent energy $E_{\rm iso}$ assumes $z=0.1384$ and the Planck18 cosmology. The quantities $\dot{\nu}$ and $\Delta\nu_{\rm d}$ denote the drift rate and spectral decorrelation bandwidth, respectively.}
\end{splitdeluxetable*}

\subsection{Polarization}
\label{subsec:polarization}

The three bursts give mutually consistent RM measurements. The Appendix shows their RM-synthesis curves. Burst 3, which has the highest polarized S/N, gives \({\rm RM}=-45.5\pm0.2~{\rm rad~m^{-2}}\). The CHIME/FRB discovery report gave preliminary values of \(|{\rm RM}|=6.9\pm0.5\) and \(3.5\pm0.5~{\rm rad~m^{-2}}\) for two bursts detected in 2024 February and June, with the sign undetermined because of cable-delay effects \citep{2024ATel16670....1S}. If these values are accurate, the RM of FRB~20240209A changed by \(\gtrsim40~{\rm rad~m^{-2}}\) within months, as observed for some other repeaters \citep{2025SCPMA..6889511F}. At the position of FRB~20240209A, the Galactic Faraday-sky reconstruction of \citet{2022A&A...657A..43H} gives a Milky Way foreground of \({\rm RM}_{\rm MW}=-20\pm9~{\rm rad~m^{-2}}\). The observer-frame extragalactic RM is therefore \({\rm RM}_{\rm ex}\approx-25\pm9~{\rm rad~m^{-2}}\). The host-frame RM is \({\rm RM}_{\rm host}=(1+z)^2({\rm RM}-{\rm RM}_{\rm MW}-{\rm RM}_{\rm IGM})\). We drew the intergalactic contribution \({\rm RM}_{\rm IGM}\) from a Gaussian distribution with zero mean and a standard deviation of \(6~{\rm rad~m^{-2}}\) \citep{2010MNRAS.409L..99S} and propagated the uncertainties in RM and \({\rm RM}_{\rm MW}\) with \(10^6\) Monte Carlo draws, following \citet{2025ApJ...982..146P}. This gives \({\rm RM}_{\rm host}=-33\pm14~{\rm rad~m^{-2}}\) (68\% interval). This low host-frame RM is consistent with the expectation of \citet{2025ApJ...979L..21S} based on the DM budget.

\begin{figure*}[htb!]
    \centering
    \includegraphics[width=0.75\textwidth]{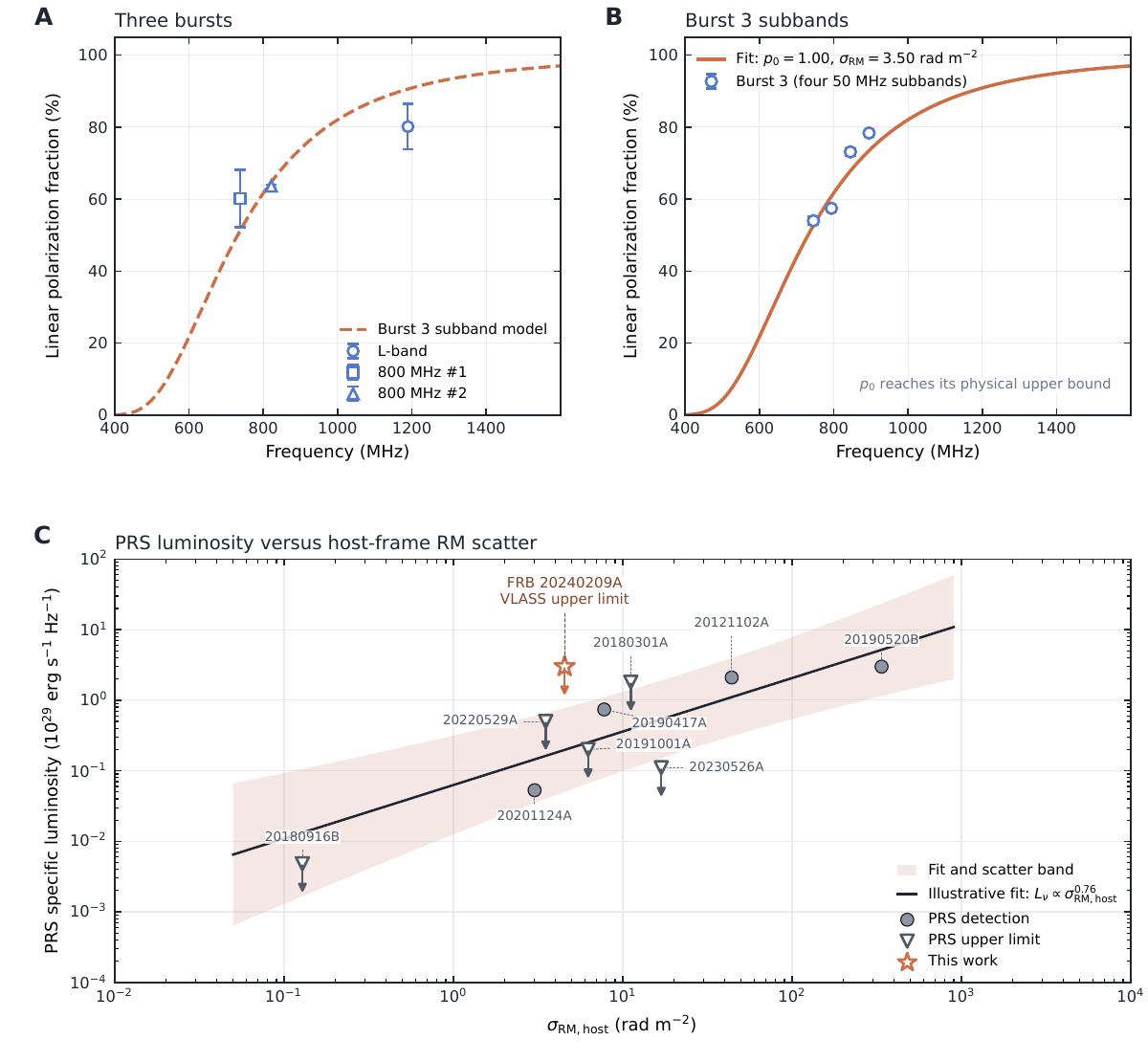}
    \caption{Frequency-dependent linear polarization and the relation between persistent radio source (PRS) luminosity and RM scatter. \textbf{(A)} Linear polarization fractions of the three bursts (blue symbols) and the RM-scattering model from panel B (orange dashed line). \textbf{(B)} Linear polarization fractions of Burst 3 in four 50 MHz subbands (blue open circles) and the fit with \(p_0\) free (orange line). \textbf{(C)} PRS specific luminosity as a function of host-frame \(\sigma_{\rm RM}\) for FRBs. Filled circles denote PRS detections. Open inverted triangles with arrows denote upper limits. The orange star and arrow mark the upper limit for FRB~20240209A. The black line is an illustrative unweighted log-space fit to the four detections. The shaded region combines the formal fit covariance and residual scatter. The upper limits are displayed and excluded from the fit. Published PRS luminosities measured at different frequencies are compared using the flat-spectrum convention of \citet{2026MNRAS.545f1997U}. The \(\sigma_{\rm RM}\) measurements are from \citet{2022Sci...375.1266F}, \citet{2024MNRAS.527.4285U}, \citet{2026MNRAS.545f1997U}, and \citet{2026ApJ..1000L..53P}. The PRS luminosities and upper limits are from \citet{2017Natur.541...58C,2020Natur.577..190M,2020ApJ...901L..20B,2022AJ....163...69B,2022Natur.606..873N,2023ApJ...958L..19B,2024Natur.632.1014B,2025ApJ...979L..21S,2026ApJ...996L..16M,2026MNRAS.545f1997U} and \citet{2026ApJ..1000L..53P}.}
    \label{fig:sigmaRM}
\end{figure*}

The measured linear polarization fractions are \(L/I=80\%\pm6\%\) for the L-band burst and \(60\%\pm8\%\) and \(63.6\%\pm0.4\%\) for the two 800 MHz bursts (Figure~\ref{fig:sigmaRM}(A)). The higher linear polarization fraction of the L-band burst relative to the two 800 MHz bursts suggests a tentative decrease of \(L/I\) toward lower frequencies. We fitted these measurements with the RM-scattering model as an exploratory description of this trend \citep{1966MNRAS.133...67B,2022Sci...375.1266F},
\begin{equation}
    \frac{L(\nu)}{I(\nu)}=p_0\exp\left(-2\sigma_{\rm RM}^2\lambda^4\right),
    \label{eq:sigma_rm}
\end{equation}
where \(p_0\) is the intrinsic linear polarization fraction and \(\sigma_{\rm RM}\) is the observer-frame RM scatter. With \(p_0=1\), the fit to the three burst-integrated measurements gives \(\sigma_{\rm RM}=3.57\pm0.21~{\rm rad~m^{-2}}\). This estimate is tentative because repeaters can show substantial burst-to-burst variations in \(L/I\) \citep{2025ApJ...982..154N}, and two of these three measurements have large uncertainties. A more robust assessment of frequency-dependent depolarization comes from the subband measurements within the exceptionally bright and broadband Burst~3, which we discuss in Section~\ref{sec:discussion_rm}.

\subsection{Candidate submillisecond spacing}
\label{subsec:qpo}

\begin{figure*}[htb!]
    \centering
    \includegraphics[width=\textwidth]{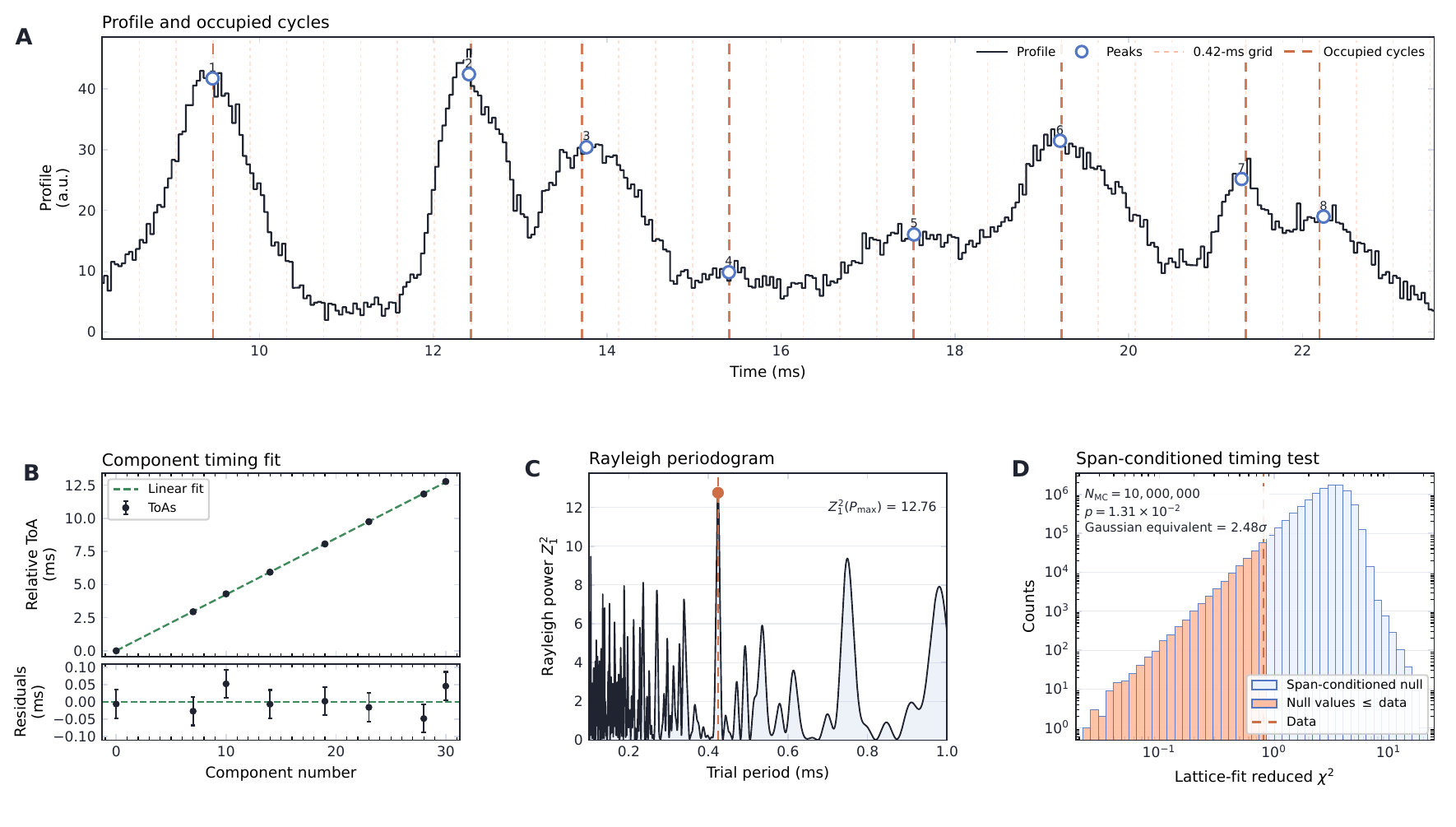}
    \caption{Timing diagnostics for the eight substructures of Burst 3. \textbf{(A)} Dedispersed total-intensity profile. The blue open circles mark the selected peaks. The light and dark orange dashed lines mark the \(0.42~{\rm ms}\) time grid and its occupied cycles, respectively. \textbf{(B)} Relative ToAs as a function of integer component number in the upper panel and their fit residuals in the lower panel. Black points show the measured ToAs with one-native-sample uncertainties. The green dashed lines show the linear fit and zero residual. \textbf{(C)} Rayleigh power \(Z^2_1\) as a function of trial period. The orange dashed line and circle mark the maximum at \(0.42~{\rm ms}\). \textbf{(D)} Distribution of the minimum reduced \(\chi^2\) from the span-conditioned shifted-exponential simulations. The blue histogram shows the null distribution, the orange bars mark values at or below the observation, and the orange dashed line marks the observed value.}
    \label{fig:qpo}
\end{figure*}

Burst 3 shows eight resolved intensity components whose peaks appear to recur on a common submillisecond timescale (Figure~\ref{fig:qpo}). To suppress noise fluctuations around the component peaks, we applied a third-order Savitzky--Golay filter \citep{1964AnaCh..36.1627S} to the profile and took the eight local maxima of the smoothed profile as the component times of arrival (ToAs). We searched for periods from \(0.1\) to \(1.0~{\rm ms}\) with a step of \(10^{-4}~{\rm ms}\). For each trial period \(P\), we converted every ToA \(t_i\) to a phase \(\phi_i(P)\) and calculated the Rayleigh \(Z^2_1\) statistic \citep{1983A&A...128..245B},

\begin{equation}
    \begin{aligned}
        \phi_i(P) &= 2\pi\,\operatorname{frac}\left(\frac{t_i-t_0}{P}\right), \\
        Z^2_1(P) &= \frac{2}{N}\left[\left(\sum_{i=1}^{N}\cos\phi_i\right)^2+\left(\sum_{i=1}^{N}\sin\phi_i\right)^2\right].
    \end{aligned}
    \label{eq:rayleigh}
\end{equation}

Here, \(N=8\) is the number of measured components, and \(t_0\) is a reference time. The Rayleigh statistic measures the concentration of the unbinned arrival phases, with larger values indicating stronger phase alignment. The periodogram reaches its maximum, \(Z^2_1=12.76\), at \(P\simeq0.42~{\rm ms}\) (Figure~\ref{fig:qpo}(C)).

To characterize how closely the component ToAs follow the \(0.42~{\rm ms}\) candidate period, we applied the timing analysis of \citet{2023A&A...678A.149P}. We allowed cycles without a detected component and searched for the integer component numbers \(n_i\) that minimized the residuals of the linear timing relation

\begin{equation}
    t_i=T_0+n_iP,
    \label{eq:component_timing}
\end{equation}

where \(P\) is the fitted period and \(T_0\) is the reference ToA. The best-fitting sequence is \(n_i=(0,7,10,14,19,23,28,30)\), consistent with the \(0.42~{\rm ms}\) period found above. Figure~\ref{fig:qpo}(B) shows the timing fit and its residuals. We adopted one native time sample, \(0.04096~{\rm ms}\), as the uncertainty of each ToA, giving \(\chi^2_{\nu,\rm obs}=0.808\).

\section{Discussion}
\label{sec:discussion}

\subsection{Implications of the candidate spacing}
\label{sec:discussion_qpo}

To assess the significance of the candidate spacing, we generated \(10^7\) simulated eight-component sequences. Following \citet{2023A&A...678A.149P}, their adjacent intervals were drawn from the shifted-exponential law

\begin{equation}
    d_j=\bar{d}\left[\eta-(1-\eta)\ln(1-u_j)\right], \qquad u_j\sim\mathcal{U}(0,1),
    \label{eq:shifted_exponential}
\end{equation}

where \(\bar{d}\) is the mean interval and \(\eta=0.2\). We multiplied the intervals in each realization by a common scale factor to match the observed first-to-last-component span, repeated the same component-number search and timing fit, and recorded the minimum reduced \(\chi^2\). Among the \(10^7\) simulations, 131,422 gave \(\chi^2_{\nu,\rm min}\leq\chi^2_{\nu,\rm obs}\). This fraction corresponds to \(p=0.01314\) and a two-sided Gaussian-equivalent significance of \(2.48\sigma\) for the \(0.42~{\rm ms}\) spacing under this span-conditioned timing model.

Near-regular component spacing has been reported in other complex FRB profiles with similarly modest significance. The published analysis of FRB~20201020A gives \(0.411\pm0.006~{\rm ms}\) at \(2.4\sigma\) under a shifted-Poisson model \citep{2023A&A...678A.149P}, and FRB~20230708A shows a \(7.267~{\rm ms}\) ACF feature at \(1.77\sigma\) \citep{2025MNRAS.536.3220D}. Searches of active repeaters have also found burst-local candidates whose spacings vary between bursts or remain coherent for only a few components \citep{2022RAA....22l4004N,2026ApJ...996....4Z,2026SCPMA..6949512Z}. The highly significant \(\sim217~{\rm ms}\) periodicity reported for FRB~20191221A \citep{2022Natur.607..256C} was subsequently identified as emission from the Galactic pulsar PSR~J0248+6021 \citep{2026arXiv260607087C}.

If the \(0.42~{\rm ms}\) spacing of Burst~3 has a physical origin, a rotational interpretation would require a spin frequency of about 2.4~kHz, which exceeds the limits set by typical neutron-star equations of state \citep{2023A&A...678A.149P}. Short-lived modulation of the emitting region offers an alternative origin. Radio pulsars and magnetars exhibit quasiperiodic microstructure on timescales much shorter than their spin periods \citep{2024NatAs...8..230K}. In Burst~3, 8 of the 31 possible component slots are occupied, leaving 23 empty. This sparse occupancy gives stochastic peak trains substantial freedom to align with the trial grid, so chance alignment remains plausible at the measured \(2.48\sigma\) significance. Repeated recovery of the same spacing in independent bursts, with more consecutively occupied slots, would strengthen the evidence for a physical clock.

\subsection{Polarization and environment}
\label{sec:discussion_rm}

The host-frame RM of FRB~20240209A, \({\rm RM}_{\rm host}=-33\pm14~{\rm rad~m^{-2}}\), is small compared with the \(|{\rm RM}|\) of \(10^4\)--\(10^5~{\rm rad~m^{-2}}\) measured for FRBs~20121102A and 20190520B \citep{2018Natur.553..182M,2023Sci...380..599A}, consistent with the low RM predicted by \citet{2025ApJ...979L..21S}. The reconstruction of \citet{2022A&A...657A..43H} does not capture Galactic RM fluctuations on subdegree scales and can misestimate the Galactic foreground by up to \(\sim40~{\rm rad~m^{-2}}\) at Galactic latitudes \(|b|\geq25^\circ\) \citep{2025ApJ...982..146P}. FRB~20240209A lies at \(b=26\fdg6\). A foreground error of this size would place \({\rm RM}_{\rm host}\) between \(-85\) and \(+19~{\rm rad~m^{-2}}\). The sign of \({\rm RM}_{\rm host}\) is therefore uncertain, and \(|{\rm RM}_{\rm host}|\) remains below \(100~{\rm rad~m^{-2}}\), far below those of FRBs~20121102A and 20190520B.

\begin{figure}[htb!]
    \centering
    \includegraphics[width=\columnwidth]{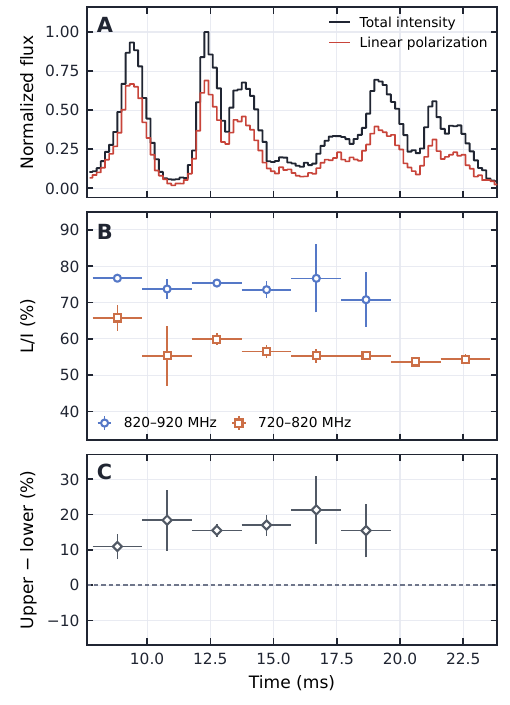}
    \caption{Linear polarization of Burst~3 in the upper and lower halves of the 800 MHz band. \textbf{(A)} Frequency-integrated profiles of total intensity (black) and linear polarization (red) over 720--920 MHz. \textbf{(B)} Linear polarization fractions in 820--920 MHz (blue open circles) and 720--820 MHz (orange open squares). \textbf{(C)} Difference between the upper-half and lower-half linear polarization fractions (gray open diamonds). The dashed line marks zero. Horizontal bars span the 1.97 ms time bins, and vertical bars show \(1\sigma\) uncertainties.}
    \label{fig:band_pol}
\end{figure}

To examine the frequency dependence of \(L/I\) independently of burst-to-burst variation in intrinsic polarization (e.g., FRB~20240114A; \citealt{2026ApJS..284...77W}), we divided Burst~3 into four 50 MHz subbands (Figure~\ref{fig:sigmaRM}(B)). Its measured \(L/I\) rises from \(54.0\%\pm1.1\%\) and \(57.4\%\pm0.5\%\) at 745 and 794 MHz to \(73.1\%\pm1.0\%\) and \(78.4\%\pm0.8\%\) at 845 and 895 MHz. The emission of Burst~3 drifts toward lower frequencies, so different components dominate the upper and lower subbands, and component-to-component changes in intrinsic polarization \citep{2020ApJ...891L..38C,2024ApJ...968...50P} could produce a similar trend. To compare the two frequency ranges at the same times, we divided Burst~3 into eight consecutive 1.97 ms bins and measured \(L/I\) separately in 720--820 and 820--920 MHz (Figure~\ref{fig:band_pol}). In each bin, we averaged the RM-corrected Stokes \(I\), \(Q\), and \(U\) within each half of the band, applied the bias correction of Equation~(\ref{eq:linear_debias}), and retained bins with \({\rm S/N}\geq10\) in that half. Figures~\ref{fig:band_pol}(B) and \ref{fig:band_pol}(C) show a consistently and significantly higher \(L/I\) in the upper half than in the lower half in all six bins where both halves meet this threshold. The frequency dependence therefore persists at fixed times across the first six components. These measurements establish a frequency-dependent polarization change within a single burst and support depolarization at lower frequencies.

The intrinsic linear polarization fraction is initially unknown, and depolarization models allow values below 100\% \citep{2026MNRAS.545f1997U}. We therefore fitted the four subband measurements with both \(p_0\) and \(\sigma_{\rm RM}\) free. The fit reaches the physical upper boundary \(p_0=1\), corresponding to a best-fitting intrinsic linear polarization fraction of 100\% within this model, and gives \(\sigma_{\rm RM}=3.50\pm0.14~{\rm rad~m^{-2}}\).

Figure~\ref{fig:sigmaRM}(C) places this RM scatter in the context of PRS luminosities, using the host-frame RM scatter \(\sigma_{\rm RM,host}=(1+z)^2\sigma_{\rm RM}\) \citep{2026ApJ..1000L..53P}. An unweighted log-space fit to the four detected PRSs, including that associated with FRB~20190417A, gives \(L_\nu\propto\sigma_{\rm RM,host}^{0.76\pm0.31}\). Luminosities measured at different frequencies are compared using a flat-spectrum convention. The tentative positive relation is consistent with a shared magnetized environment that both depolarizes the bursts and produces persistent synchrotron emission \citep{2022ApJ...928L..16Y,2026MNRAS.545f1997U}. Such environments can arise in synchrotron-emitting nebulae formed after supernovae. The diversity of compact remnants is illustrated by a long-period radio transient plausibly associated with a supernova remnant \citep{2026SciBu..71.3609L} and a radio-emitting central compact object \citep{2026NatAs..10.1345Z}. In this picture, RM scatter traces fluctuations in the magnetized plasma, and the PRS is powered by relativistic electrons in the same environment. For comparable nebular sizes, electron distributions, and fluctuation properties, the predicted dependence is approximately linear, consistent with the fitted slope \citep{2022ApJ...928L..16Y}. This small, heterogeneous sample provides an illustrative relation. At the \(\sigma_{\rm RM,host}\) of FRB~20240209A, the relation gives \(L_\nu\sim2\times10^{28}~{\rm erg~s^{-1}~Hz^{-1}}\), consistent with the VLASS upper limit \citep{2025ApJ...979L..21S}. Continuum imaging that reaches this luminosity can test whether the magnetoionic structure inferred from depolarization also powers a PRS.

\begin{figure}[htb!]
    \centering
    \includegraphics[width=\columnwidth]{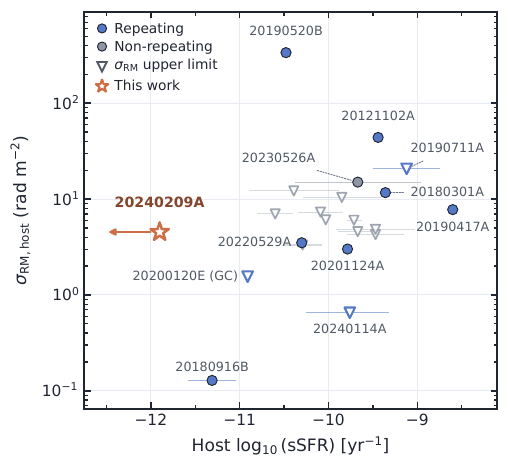}
    \caption{Host-frame RM scatter versus global host sSFR for 22 localized FRBs. Blue and gray symbols denote repeating and apparently nonrepeating sources, respectively. Filled circles denote measurements, and open inverted triangles denote upper limits. The orange star marks FRB~20240209A, and its leftward arrow indicates that the host sSFR is an upper limit.}
    \label{fig:sigma_rm_ssfr}
\end{figure}

The RM scatter quantifies the complexity of the magnetoionic environment, and \citet{2022Sci...375.1266F} associated larger \(\sigma_{\rm RM}\) values with younger FRB sources. To examine whether measurable RM scatter also occurs in FRBs from hosts with a low specific star formation rate, \({\rm sSFR}={\rm SFR}/M_*\), we compare \(\sigma_{\rm RM}\) with the global host sSFR in Figure~\ref{fig:sigma_rm_ssfr}. The comparison includes FRB~20240209A and 21 FRBs with secure host associations from the literature. Spectral depolarization analyses constrain \(\sigma_{\rm RM}\) for 20 of the literature sources \citep{2022Sci...375.1266F,2023MNRAS.526.3652K,2024MNRAS.527.4285U,2026MNRAS.545f1997U,2026ApJ..1000L..53P,2026ApJS..284...77W}. For FRB~20200120E, burst 20210423G has \(L/I=0.993\pm0.007\) in the 400--800~MHz CHIME band \citep{2025ApJ...982..154N}. Equation~(\ref{eq:sigma_rm}) then gives \(\sigma_{\rm RM}\lesssim1.5~{\rm rad~m^{-2}}\). All \(\sigma_{\rm RM}\) upper limits are at 95\% confidence. We convert \(\sigma_{\rm RM}\) to the host frame as \(\sigma_{\rm RM,host}=(1+z)^2\sigma_{\rm RM}\) \citep{2026ApJ..1000L..53P}, assuming that the dominant depolarizing screen lies at the host redshift. Host SFRs and stellar masses are taken from \citet{2022AJ....163...69B,2023ApJ...954...80G,2025PASA...42..157G,2025ApJ...989...77L,2025ApJ...979L..22E,2025ApJ...992L..35B,2026Sci...391..280L,2026ApJ...996L..16M}. For FRB~20200120E, these quantities refer to M81 as a whole.

FRB~20240209A has the lowest host sSFR in the sample. Its quiescent host has \(\log_{10}({\rm sSFR}/{\rm yr}^{-1})<-11.9\) and a mass-weighted stellar age of approximately \(11\,{\rm Gyr}\), indicating an old stellar population \citep{2025ApJ...979L..22E}. Its host-frame RM scatter, \(\sigma_{\rm RM,host}=4.5\pm0.2~{\rm rad~m^{-2}}\), is comparable to those of FRBs~20201124A and 20220529A, whose hosts have sSFRs higher by more than an order of magnitude. It also exceeds the value for FRB~20180916B and the upper limit for FRB~20240114A, both of which reside in star-forming galaxies. The large projected offset of FRB~20240209A leaves its local stellar population and the location of the depolarizing screen unresolved \citep{2025ApJ...979L..21S}, so the comparison is based on global host properties. This limitation applies more generally to the sample. For almost all FRBs, the global host sSFR may not represent the sSFR at the burst position.

Relatively complex magnetoionic environments therefore occur around repeaters in both star-forming and old, quiescent hosts. Under the interpretation of \citet{2022Sci...375.1266F} that a measurable RM scatter marks a young source, such active repeating FRBs also arise in old stellar populations. FRB~20200120E provides independent evidence for repeating FRB sources in an old stellar population through its localization to a GC in M81 \citep{2022Natur.602..585K}. Together with repeaters in young star-forming environments \citep{2021ApJ...908L..12T,2022Natur.606..873N}, these results support the picture that FRB sources may form through multiple channels \citep{2022Natur.602..585K,2025ApJ...979L..21S,2025ApJ...979L..22E}, including pathways that operate in old stellar populations.

\section{Conclusions}
\label{sec:conclusion}

We report GBT observations of FRB~20240209A at L band and 800 MHz on 2024 August 18. Our main results are summarized as follows.

\begin{itemize}
    \setlength{\itemsep}{3pt}
    \item[1.] We detected three bursts, one at L band and two at 800 MHz. Burst~3 resolves into eight components whose emission drifts toward lower frequencies at later times, with a drift rate of \(-8.5\pm0.9~{\rm MHz~ms^{-1}}\). Its spectral ACF gives a scintillation bandwidth of \(1.17\pm0.13~{\rm MHz}\) at 820 MHz.
    \item[2.] The eight components of Burst~3 give a candidate spacing of \(0.42~{\rm ms}\) with \(2.48\sigma\) significance. Only 8 of the 31 slots in the fitted grid are occupied, so chance alignment remains plausible. Recovery of the same spacing in independent bursts would test whether it reflects a physical clock.
    \item[3.] The three bursts have mutually consistent RMs, and Burst~3 gives \({\rm RM}=-45.5\pm0.2~{\rm rad~m^{-2}}\). The host-frame RM of \(-33\pm14~{\rm rad~m^{-2}}\) agrees with the low RM predicted from the DM budget.
    \item[4.] The linear polarization fraction shows a tentative decrease toward lower frequencies among the three bursts and a significant decrease across four subbands of Burst~3. The RM-scattering fit to the subbands gives \(\sigma_{\rm RM}\simeq3.5~{\rm rad~m^{-2}}\) and a best-fitting intrinsic linear polarization fraction of 100\%. The VLASS upper limit on a PRS is consistent with the tentative relation between RM scatter and PRS luminosity. The host-frame RM scatter, \(\sigma_{\rm RM,host}=4.5\pm0.2~{\rm rad~m^{-2}}\), is comparable to those of FRBs~20201124A and 20220529A and well below those of FRBs~20121102A and 20190520B.
\end{itemize}

The RM scatter of FRB~20240209A reveals a relatively complex magnetoionic environment around a repeater in an old, quiescent host. Such environments therefore occur around repeaters in both star-forming and quiescent hosts. Together with the globular-cluster localization of FRB~20200120E, these results support the picture that FRB sources may form through multiple channels, including pathways that operate in old stellar populations. Repeated wideband polarimetry can track the evolution of this magnetoionic environment, and deeper continuum imaging can test for a PRS at the luminosity predicted from the RM scatter.

\begin{acknowledgments}
    This work is supported by the National Natural Science Foundation of China under grants 12588202, 12603083 and 12522305. Y.-K.Z. is supported by the Postdoctoral Fellowship Program and China Postdoctoral Science Foundation under grant BX20250158 and by the National Natural Science Foundation of China under grant 12573096. D.L. acknowledges support from the New Cornerstone Science Foundation. J.-T.X. acknowledges support from the Sichuan Science and Technology Program under grant 2026NSFSC0743. P.W. acknowledges support from the CAS Youth Interdisciplinary Team, the Youth Innovation Promotion Association CAS (id. 2021055), and the Specialized Research Fund for State Key Laboratory of Radio Astronomy and Technology.
\end{acknowledgments}

\facilities{GBT}

\software{PSRCHIVE \citep{2004PASA...21..302H}, DM\_phase \citep{2019ascl.soft10004S}, DRAFTS \citep{2025ApJS..276...20Z}}

\appendix
\setcounter{figure}{0}
\renewcommand{\thefigure}{A\arabic{figure}}

\section*{RM-synthesis Curves}
\label{app:fdf}

\begin{figure}[htb!]
    \centering
    \includegraphics[width=\textwidth]{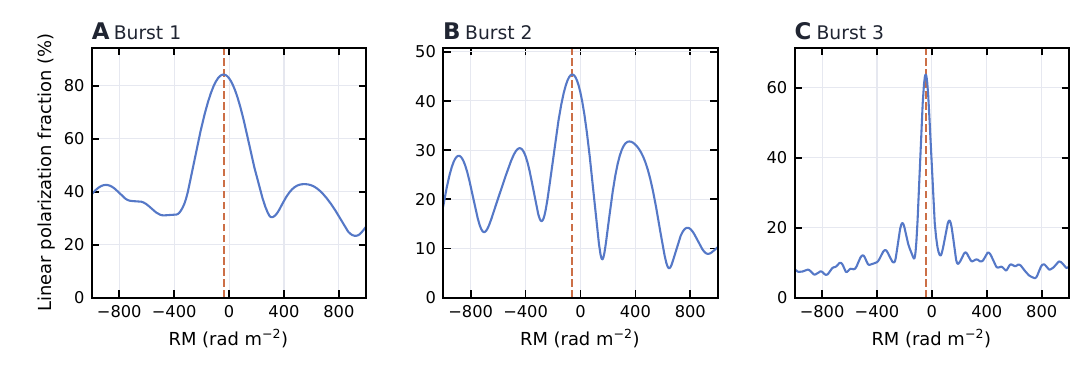}
    \caption{Linear polarization fraction as a function of trial RM from RM synthesis. \textbf{(A)}--\textbf{(C)} Bursts 1, 2, and 3, respectively. The orange dashed lines mark the RMs in Table~\ref{tab:burst_properties}.}
    \label{fig:fdf}
\end{figure}

Figure~\ref{fig:fdf} shows the linear polarization fraction as a function of trial RM from the RM synthesis described in Section~\ref{sec:data}. This quantity is the amplitude of the Faraday dispersion function of each time sample, summed over the burst and divided by the summed total intensity \citep{1966MNRAS.133...67B,2005A&A...441.1217B}. The curves peak at the RMs listed in Table~\ref{tab:burst_properties}. To estimate the significance of each peak, we applied the same procedure to 5000 sets of randomly selected off-pulse time samples, each containing as many samples as the burst, and evaluated them at the peak RM. The peaks exceed the mean of these noise realizations by 16, 5.8, and 400 times their standard deviation for Bursts 1, 2, and 3, respectively.

\bibliography{references}
\bibliographystyle{aasjournalv7.1}

\end{document}